\documentclass[final]{elsart5p}
\usepackage{graphicx,amssymb}
\usepackage{url}
\journal{Computational Material Science}
\usepackage{dcolumn}
\usepackage[square,numbers,sort]{natbib}
 
\usepackage{ifpdf}
\ifpdf
\usepackage[%
  pdftitle={Large scale ab initio calculations based on three levels of parallelization},%
  pdfauthor={Francois Bottin, Stephane Leroux, Andrew Knyazev, and Gilles Zerah},%
  pdfsubject={Large scale ab initio calculations coupling three levels of parallelization},%
  pdfkeywords={},%
  pdfstartview=FitH,%
  bookmarks=true,%
  bookmarksopen=true,%
  breaklinks=true,%
  colorlinks=true,%
  linkcolor=blue,anchorcolor=blue,%
  citecolor=blue,filecolor=blue,%
  menucolor=blue,pagecolor=blue,%
  urlcolor=blue]{hyperref}
\else
\usepackage[%
  breaklinks=true,%
  colorlinks=true,%
  linkcolor=blue,anchorcolor=blue,%
  citecolor=blue,filecolor=blue,%
  menucolor=blue,pagecolor=blue,%
  urlcolor=blue]{hyperref}
\fi
\usepackage{hypernat}

\newcommand{\bra}[1]{\big\langle#1\big|}
\newcommand{\ket}[1]{\big|#1\big\rangle}

\newcommand{\gras}[1]{\mathbf{#1}}

\begin{document}
\begin{frontmatter}
\title{\bf Large scale ab initio calculations based on three levels of parallelization}
\author[DPTA,LRC]{Fran\c{c}ois Bottin\corauthref{cor}}, 
\corauth[cor]{Corresponding author. Tel.: +33 16926 4000}
\ead{Francois.Bottin@cea.fr}
\author[DPTA]{St\'ephane Leroux},
\ead{stephane.leroux@cea.fr}
\author[Colorado]{Andrew Knyazev\thanksref{labelAK}}, 
\thanks[labelAK]{Partially supported by the NSF award 
DMS-0612751.}
\ead{andrew.knyazev@cudenver.edu}
\ead[url]{http://math.cudenver.edu/\~{}aknyazev/}
\author[DPTA]{Gilles Z\'erah}
\ead{gilles.zerah@cea.fr}
\address[DPTA]{D\'epartement de Physique Th\'eorique et Appliqu\'ee, 
CEA/DAM Ile-de-France, BP 12, 91680 Bruy\`eres-le-Ch\^atel Cedex, France}
\address[LRC]{LRC - Centre de Math\'ematiques et de Leurs Applications, CNRS (UMR 8536)\\
ENS Cachan, 61, Avenue du Pr\'esident Wilson, Cachan Cedex, France}
\address[Colorado]{Department of Mathematical Sciences,
University of Colorado at Denver, 
P.O. Box 173364, Campus Box 170, Denver, CO 80217-3364.}
{\small Received May 11, 2007; received in the revised form July 11, 2007; accepted July 22, 2007}
\begin{abstract}
We suggest and implement a parallelization scheme based on an efficient multiband 
eigenvalue solver, called the locally optimal block preconditioned conjugate gradient 
(\textsc{lobpcg}) method, and using an optimized three-dimensional (3D) fast Fourier 
transform (FFT) in the {\it ab initio} plane-wave code \textsc{abinit}. 
In addition to the standard data partitioning over processors corresponding to 
different \textbf{k}-points, we introduce data partitioning with respect 
to blocks of bands as well as spatial partitioning in the Fourier space of coefficients 
over the plane waves basis set used in \textsc{abinit}.  
This  \textbf{k}-points-multiband-FFT parallelization avoids any collective 
communications on the whole set of processors 
relying instead on one-dimensional communications only. For a single \textbf{k}-point, 
super-linear scaling is achieved for up to 100 processors due to an extensive use 
of hardware optimized \textsc{blas}, \textsc{lapack} and \textsc{scalapack} routines, 
mainly in the \textsc{lobpcg} routine.
We observe good performance up to 200 processors.  
With 10 \textbf{k}-points our three-way data partitioning results in 
linear scaling up to 1000 processors for a practical system used for testing.\\ 
\noindent\copyright{2007}{ 
Bottin, Leroux, Knyazev, Z\'erah.
All rights reserved.}
\end{abstract}
\end{frontmatter}
\section{Introduction}
The density functional theory (DFT)~\cite{Hohenberg_PR136_1964} 
and the Kohn-Sham (KS) equations~\cite{Kohn_PR140_1965} are 
efficient techniques reducing the costs of solving the  
original Schr\"odinger equations without sacrificing the accuracy in many practically important cases. 
However, numerical modeling of such properties 
as large surface reconstructions
~\cite{Brommer_PRL68_1992,DieboldPRLzno,Kresse_PRB68_2003,Segeva_SS601_2007} and melting
~\cite{Ogitsu_PRL91_2003,Sugino_PRL74_1995,Wang_PRL95_2005,Cricchio_PRB73_2006}, 
which involves molecular dynamics and systems with hundreds atoms, remains too 
time consuming for single-processor computers even in the DFT-KS framework. 
Parallelization, i.e.,\ 
software implementation suitable for computers with many processors, of electronic 
structure codes has become one of the main software development tasks 
as 
tens of thousands processors are already available on massively parallel systems, 
see ~\url{http://www.top500.org}. In order to use efficiently these supercomputers, 
various {\it ab initio} codes are being deeply modified
~\cite{Vadali_JCC25_2004,Hutter_CPC6_2005,Hutter_PC31_2005,Skylaris_PSS243_2006,Skylaris_JCP122_2005}. 

Efficient eigensolvers for iterative diagonalization of the Hamiltonian matrix~\cite{PayneRMP92} 
are necessary for large systems. 
In the framework of {\it ab initio} calculations, the most commonly used diagonalization 
method is the band-by-band conjugate gradient (\textsc{cg}) algorithm proposed by Teter 
{\it et al.}~\cite{Teter_PRB40_1989} for direct minimization of the total energy. Thereafter, 
it has been modified by Bylander {\it et al.}~\cite{Bylander_PRB42_1990} to fit the iterative 
diagonalization framework. This algorithm is fast and robust for small matrices, but requires 
explicit orthogonalization of residual vectors to already resolved bands. For larger 
matrices, the residual minimization method---a direct inversion in the iterative subspace 
(RMM-DIIS)~\cite{Kresse_PRB54_1996,Kresse_CMS6_1996}---is more efficient. This latter scheme 
based on the original work of Pulay~\cite{Pulay_CPL73_1980} and modified by Wood and 
Zunger~\cite{Wood_JPA18_1985} avoids orthogonalizations. 

The popular preconditioned block Davidson scheme~\cite{Davidson_JCP17_1975} is reported to 
be overtaken by the preconditioned \textsc{cg} (\textsc{pcg}) method for small and by the 
RMM-DIIS for large systems~\cite{Kresse_CMS6_1996}. Other widely used methods for large 
systems are the Lanczos algorithm~\cite{Lanczos_JRNBS45_1950} with 
partial reorthogonalization and the Chebyshev-filtered subspace iterations~\cite{Zhou_PRE74_2006} 
with one diagonalization at the first iteration, neither utilizes preconditioning.  

In solid state physics, it is convenient to expand wave-functions, densities and potentials 
over a plane waves (PW) basis set, where the wave-functions identify themselves with Bloch's 
functions and  periodic boundary conditions are naturally treated. Moreover, the dependency 
of the total energy functional on the PW basis set is variational in this case. 
Long-range potentials, which are hard to compute in real space, can be easily evaluated in 
reciprocal (Fourier) space of the PW expansion coefficients. Depending on the computational costs, 
quantities are computed in real or reciprocal space---and we go from one space to the other 
using backward and forward three-dimensional (3D) fast Fourier transformations (FFTs) .

As iterative eigensolvers and FFTs dominate in the overall computational costs of PW-based 
electronic structure codes, such as \textsc{abinit}~\cite{abinit,abinitart}, 
efficient coordinated parallelization is necessary.  
In the present paper, we suggest and implement a parallelization scheme based on an efficient multiband 
eigenvalue solver, called the locally optimal block preconditioned conjugate gradient 
(\textsc{lobpcg}) method, and using optimized 3D FFTs in the {\it ab initio} PW code \textsc{abinit}. 
In addition to the standard data partitioning over processors corresponding to 
different \textbf{k}-points, we introduce data partitioning with respect 
to blocks of bands as well as spatial partitioning in the Fourier space of PW coefficients.  

The eigensolver that we use 
in this work is the \textsc{lobpcg} method proposed by Knyazev~\cite{k90,k98,k99,Knyazev_JSC23_2001}. 
Our choice of the \textsc{lobpcg} is determined by its simplicity, effectiveness, ease of 
parallelization, acceptance in other application areas, and public availability 
of codes~\cite{blopex07}. In the framework of {\it ab initio} self-consistent (SC) 
calculations, the use of  \textsc{lobpcg}  is apparently novel, see also \cite{ymw06} 
where the \textsc{lobpcg} is adapted for the total energy minimization.

For FFTs, we use the 3D FFT implemented by Goedecker {\it et al.}~\cite{Goedecker_CPC_2003}, 
which has previously demonstrated its efficiency on massively parallel supercomputers. 
This algorithm is cache optimized, minimizes the number of collective communications, 
and allows FFTs with zero padding.

In section \ref{sec:SC} we describe the 
flow chart and bottlenecks of self-consistent calculations to motivate our work on parallelization. 
We introduce the \textsc{lobpcg} eigensolver and compare it to the standard CG eigensolver is section 
\ref{sec:lobpcg}. Our multiband-FFT parallelization and the optimizations performed are described in 
section \ref{sec:bfft}. In section \ref{sec:results} we demonstrate our numerical results 
and report scalability of our two-level (multiband-FFT) and 
three-level  (\textbf{k}-points-multiband-FFT) parallelization a large-size system. 
\section{Motivation to parallelize the self-consistent calculations
\label{sec:SC}}
\subsection{The self-consistent loop\label{ssec:SC}}
In order to keep the problem setup general, we use the term in the right-hand side of the 
KS equations, the so-called overlap operator $\mathcal{O}$, which comes from the non-orthogonality 
of the wave-functions. It appears for ultra-soft pseudo-potentials (USPP)~\cite{Vanderbilt_PRB41_1990} 
or the projector augmented-wave (PAW) method~\cite{Blochl_PRB50_1994} and is reduced to the identity 
in norm-conserving calculations. 
The minimum of the total energy functional is calculated by carrying on a SC 
determination of the density and potential in the SC loop. In the following flow-chart, we display 
schematically the SC loop with its various components in the framework of PAW calculations in \textsc{abinit}:
\begin{displaymath}
 \begin{array}{ccc}
\rm
  \tilde{\Psi}_{nk}(\gras{r})=\sum_{\gras G} c_{nk}({\gras G}) e^{{i}({\gras k}+{\gras G}).{\gras r}}&&\\
  \Big\downarrow&&\\
\rm
  [\tilde n+\hat n](\gras{r}) \mathrm{\;\;\; and \;\;\;} \rho_{ij}&\longleftarrow&
\rm 
  \left\{c_{nk}({\gras G});\epsilon_{nk}\right\}\\
  \Big\downarrow&&\Big\uparrow\\
\rm
  v_{loc}(\gras{r}) \mathrm{\;\;\; and \;\;\;} v_{nl}(\gras{r})&&
\rm
  \mid\tilde{\mathcal{H}}-\epsilon_{n}\mathcal{O}\mid=0\\
  \Big\downarrow&&\Big\uparrow\\
  \multicolumn{3}{c}{\rm \bra{e^{i({\gras k}+{\gras G}).{\gras r}}}\tilde{\mathcal{H}}\ket{\tilde{\Psi}_{nk}}=\epsilon_{nk}
  \bra{e^{i({\gras k}+{\gras G}).{\gras r}}}\mathcal{O}\ket{\tilde{\Psi}_{nk}}}\\
 \end{array}
\end{displaymath}
The initiation step is that the wave-function of the system $\Psi_{\rm nk}(\gras{r})$,
where n and k are the band and the \textbf{k}-point indexes, respectively, 
is expanded over the PW basis set with coefficients $\mathrm{c_{nk}}({\gras G})$  
and vectors $\gras{G}$ in the reciprocal space. Now the SC loop starts:
First, the wave-function is used to compute the pseudized charge density 
$\mathrm{\tilde n}(\gras{r})$ as well as PAW specific quantities: the compensation 
charge $\mathrm{\hat n}(\gras{r})$, needed to recover the correct multipolar development 
of the real charge density $\mathrm{n}(\gras{r})$, and the matrix density $\mathrm{\rho_{ij}}$,
which defines the occupancies of the partial waves within the PAW spheres.
Second, these densities are used to compute the local $\mathrm{v_{loc}}$ and non-local 
$\mathrm{v_{nl}}$ parts of the potential. Third, as the current Hamiltonian is determined, 
the linear (generalized if the overlap operator $\mathcal{O}$ is present) 
eigenvalue problem projected over PW
is now solved iteratively to obtain 
new approximate wave-functions PW coefficients $\mathrm{c_{nk}}({\gras G})$ corresponding 
to the minimal states. The next step is the subspace diagonalization involving all bands 
to improve the accuracy of approximate eigenvectors from the previous step.   
Finally, the input and output densities (or potentials) are mixed. The SC loop 
stops when the error is lower than a given tolerance.
\subsection{Bottlenecks as motivation for parallelization\label{ssec:bot}}
As the size of the system increases, various parts of the SC loop become very time 
consuming. Parallelization of the most computationally expensive parts 
is expected to remove or at least widen the existing bottlenecks. 
We focus here on parallel algorithms to address the following tasks, in the order of importance:  
(i) iterative solution of linear generalized eigenvalue problems (step 3 of the SC loop);
(ii) calculations of eigenvalues and eigenvectors in the subspace (step 4 of the SC loop); 
(iii) FFT routines where the local potential is applied to wave-functions or 
where the density is recomputed 
(necessary to apply the Hamiltonian in the eigensolver in step 3 of the SC loop); and 
(iv) computation of three non-local-like terms: the non-local 
part of the potential v$_{\rm nl}$, the overlap operator $\mathcal{O}$, and the 
$\rho_{\rm ij}$ matrix (step 2 of the SC loop).

A brief review of our approaches to these problems follows:  
To solve approximately eigenproblems (i) in parallel, we use multiband (block) 
eigenvalue solvers, where energy minimization is iteratively performed in 
parallel for sets of bands combined into blocks. 
\textsc{scalapack} is used to solve (ii). 
3D FFTs are well suited to perform FFT in parallel to address (iii),  
distributing the calculations over processors by splitting the PW coefficients. 
Computation of non-local-like terms (iv) can be efficiently parallelized in the reciprocal space. 
\section{Multiband eigenvalue solver\label{sec:lobpcg}}
\subsection{Locally optimal eigenvalue solver\label{ssec:lopcg}}
Our eigensolver, the \textsc{lobpcg} method, is  
a \textsc{cg}-like algorithm, which in its single-band version~\cite{k90} 
calls the Rayleigh-Ritz procedure 
to minimize the eigenvalue $\epsilon$ within the 3D subspace $\Xi$ spanned by 
$\left\{\psi^{\rm (i)},\nabla\epsilon(\psi^{\rm (i)}),\psi^{\rm (i-1)}\right\},$ 
where the index $i$ represents the iteration step number during the minimization and
$\nabla\epsilon(\psi^{\rm (i)})$ is the gradient of the Rayleigh quotient 
$\epsilon(\psi^{\rm (i)})$ given by
\begin{equation}
\epsilon(\psi)=\frac{\bra{\psi}\mathcal{H}\ket{\psi}}{\bra{\psi}\mathcal{O}\ket{\psi}}, \quad
\nabla\epsilon(\psi)=2 \frac{\mathcal{H}\psi-\epsilon(\psi)\mathcal{O}\psi}{\bra{\psi}\mathcal{O}\ket{\psi}}.
\end{equation}
Since scaling of basis vectors is irrelevant, we replace $\nabla\epsilon(\psi^{\rm (i)})$ with  
$\mathrm{r}^{\rm (i)}=\mathcal{H}\psi^{\rm (i)}-\epsilon^{\rm (i)}\mathcal{O}\psi^{\rm (i)}$, 
so the new Ritz vector is 
$
\psi^{\rm (i+1)}= \delta^{\rm (i)}\psi^{\rm (i)}+\lambda^{\rm (i)}\mathrm{r}^{\rm (i)}+\gamma^{\rm (i)}\psi^{\rm (i-1)}, 
$
with the scalar coefficients $\delta^{\rm (i)}$, $\lambda^{\rm (i)}$ and $\gamma^{\rm (i)}$ being 
obtained by the Rayleigh-Ritz minimization on the subspace $\Xi$. The use of the variational 
principal to compute the iterative parameters justifies the name ``locally optimal'' 
and makes this method distinctive, as in other \textsc{pcg} methods, e.g., 
in~\cite{Teter_PRB40_1989}, formulas for iterative parameters are explicit. 

To avoid the instability in the Rayleigh-Ritz procedure 
arising when $\psi^{\rm (i)}$ becomes close to $\psi^{\rm (i-1)}$ 
as the method converges toward the minimum, a new basis vector 
$\mathrm{p}^{\rm (i+1)}= \lambda^{\rm (i)}\mathrm{r}^{\rm (i)}+\gamma^{\rm (i)}\mathrm{p}^{\rm (i)}=
\psi^{\rm (i+1)}-\delta^{\rm (i)}\psi^{\rm (i)},
$
where $\mathrm{p}^{\rm (0)}=0$,
has been introduced~\cite{Knyazev_JSC23_2001}.  Replacing $\psi^{\rm (i-1)}$ 
with  $\mathrm{p}^{\rm (i)}$ in the basis of the minimization subspace $\Xi$ 
does not change it but gives a more numerically stable algorithm: 
\begin{equation}\label{lobpcg_no_prec}
\psi^{\rm (i+1)}= \delta^{\rm (i)}\psi^{\rm (i)}+\lambda^{\rm (i)}\mathrm{r}^{\rm (i)}+\gamma^{\rm (i)}\mathrm{p}^{\rm (i)}. 
\end{equation}
In order to accelerate the convergence and thus to improve the performance, 
a PW optimized preconditioner K is introduced and the 
preconditioned residual vectors 
$
\mathrm{w}^{\rm (i)}=\mathrm{K}(\mathrm{r}^{\rm (i)})=\mathrm{K}(\mathcal{H}\psi^{\rm (i)}-\epsilon^{\rm (i)}\mathcal{O}\psi^{\rm (i)})
$
replace $\mathrm{r}^{\rm (i)}$ in (\ref{lobpcg_no_prec}). 
In the next subsection we describe a multiband, or block, version of the  
 \textsc{lobpcg} method, which is especially suitable for parallel computations.
\subsection{Locally optimal multiband solver\label{ssec:lobpcg}}
The single-band method  (\ref{lobpcg_no_prec}) can be used  
in the standard band-by-band mode, where 
the currently iterated band is constrained to be $\mathcal{O}$-orthogonal to the previously computed ones. 
Alternatively, method   (\ref{lobpcg_no_prec}) 
can be easily generalized~\cite{k98,k99,Knyazev_JSC23_2001} to iterate a block of $m>1$ vectors 
to compute $m$ bands simultaneously. 
It requires diagonalization of a matrix of the size $3m$ on every iteration. 
 
In large-scale {\it ab initio} calculations the total number $M$ of bands 
is too big to fit all $M$ bands into one block $m=M$ as the 
computational costs of the  Rayleigh-Ritz procedure on the 
$3m$ dimensional subspace $\Xi$ becomes prohibitive. Instead, 
$M$ eigenvectors and eigenvalues 
are split into blocks $\Psi=\{\psi_{\rm 1},\ldots,\psi_{\rm m}\}$
and $\Upsilon={\rm diag}\{\epsilon_{1},\ldots,\epsilon_{\rm m}\}$ of the size $m<M$. 
The other quantities involved
in this algorithm are also defined by blocks; the small letters are
then replaced by capital letters, e.g.,\ 
the scalars  $\delta^{\rm (i)}$, $\lambda^{\rm (i)}$ and $\gamma^{\rm (i)}$ 
are replaced with $m$-by-$m$ matrices 
$\Delta^{\rm (i)}$, $\Lambda^{\rm (i)}$ and $\Gamma^{\rm (i)}$.  
The algorithm for the block-vector $\Psi$ of $m$ bands 
that we use here is the same as that in the \textsc{lobpcg} code 
in the general purpose library \textsc{blopex}~\cite{blopex07}:\\
\begin{algorithm}[h!]
{\bf Require}: Let $\Psi^{(0)}$ be a block-vector 
and $K$ be a preconditioner; 
the $\mathrm{P}^{\rm (0)}$ is initialized to 0. \\
{\bf for} i=1,$\ldots$,\texttt{nline} {\bf do}
\begin{enumerate}
\item $\Upsilon^{\rm (i)}=\Upsilon(\Psi^{\rm (i)})$
\item $\mathrm{R}^{\rm (i)}=\mathcal{H}\Psi^{\rm (i)}-\mathcal{O}\Psi^{\rm (i)}\Upsilon^{\rm (i)}$
\item $\mathrm{W}^{\rm (i)}=\mathrm{K}(\mathrm{R}^{\rm (i)}$)
\item We apply the Rayleigh-Ritz method within the subspace $\Xi$ spanned by the columns of 
$\Psi^{\rm (i)},\mathrm{W}^{\rm (i)}$ and $\mathrm{P}^{\rm (i)}$
to find
$\Psi^{\rm (i+1)}=\Psi^{\rm (i)}\Delta^{\rm (i)}+\mathrm{W}^{\rm (i)}\Lambda^{\rm (i)}+\mathrm{P}^{\rm (i)}\Gamma^{\rm (i)}$ corresponding to the minimal $m$ states within $\Xi$
\item $\mathrm{P}^{\rm (i+1)}=\mathrm{W}^{\rm (i)}\Lambda^{\rm (i)}+\mathrm{P}^{\rm (i)}\Gamma^{\rm (i)}$
\end{enumerate}
{\bf end for}
\caption{The locally optimal multiband solver \textsc{lobpcg}}
\end{algorithm}
\\
The \textsc{lobpcg} algorithm 
as formulated above computes only the $m$ lowest states, 
while we need to determine all $M>m$ states. So we perform \textsc{lobpcg} algorithm 
within the loop over the blocks, where the current \textsc{lobpcg} block is constrained 
to be $\mathcal{O}$-orthogonal to the previous ones.
In other words, we iterate in a multiband-by-multiband fashion. 
Typically we perform a fixed number \texttt{nline} of iterations in each sweep on the eigensolver, 
so the total number of iterations to approximate the solution to the eigenvalue problem 
with the given fixed Hamiltonian is \texttt{nline}$M/m.$

To improve the accuracy, we conclude these multiband-by-multiband iterations 
with a final Rayleigh-Ritz call for all $M$ bands, as pointed out in the description of the SC loop
in section \ref{sec:SC}, but this procedure is technically outside of the \textsc{lobpcg} 
routine in the \textsc{abinit} code.  
\subsection{ {\textsc{LOBPCG}} vs. the standard {\textsc{PCG}}\label{ssec:lobpcgt}}
The  \textsc{lobpcg} algorithm is much better supported theoretically~\cite{kn03}
and can even outperform the traditional \textsc{pcg} method~\cite{Teter_PRB40_1989} as we now demonstrate.  
We test band-by-band  \textsc{lobpcg}, full-band  \textsc{lobpcg} and \textsc{cg} methods 
for two different systems: a simple one with two atoms 
of carbon within a diamond structure and a somewhat larger and  
rather complex system with sixteen 
atoms of plutonium in the $\alpha$ monoclinic structure.  

The time per iteration is approximately the same for 
the single-band \textsc{lobpcg} and \textsc{pcg} methods for single-processor calculations, 
so the complexity of one self-consistent electronic step is roughly the same for both methods.
For the full-band  \textsc{lobpcg} the number of floating-point operations is higher than that in the 
band-by-band version because of the need to perform the Rayleigh-Ritz procedure on the 
$3m=3M$ dimensional subspace, but the extensive use of the high-level \textsc{BLAS}
for matrix-matrix operations in \textsc{lobpcg} may compensate for this increase. 
Thus, we display the variation of the total energy as a function of the
number of self-consistent electronic steps and, 
as the efficiency criterion, we use the final number of these steps 
needed for the variation of the total energy to reach the given tolerance in the SC loop.
The numbers $m$ and $M$ are called $\texttt{blocksize}$ and $\texttt{nband}$, correspondingly, in the \textsc{abinit} code, 
so we use the latter notation in our figures below. 
\begin{center}
\begin{figure}[ht]
\mbox{\includegraphics[width=8.7cm]{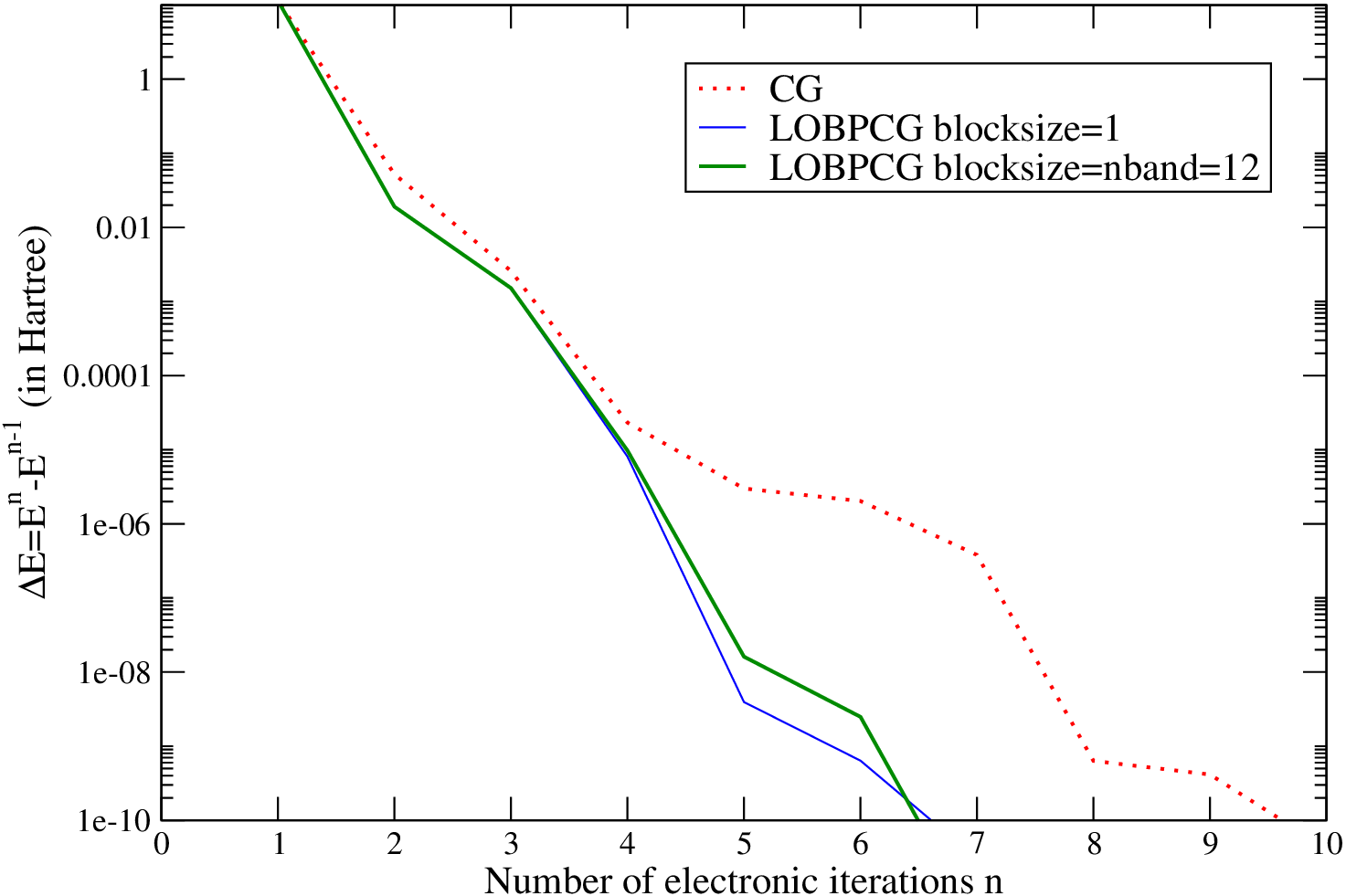}}
\caption{Total energy variation of carbon in its diamond phase as a function of the self-consistent
steps for band-by-band \textsc{pcg} and ($m=\texttt{blocksize}=1$) \textsc{lobpcg} methods, 
and full-band ($m=\texttt{blocksize}=M=\texttt{nband}=12$) \textsc{lobpcg} method. The index n stands for the 
number of electronic iterations within the SC loop, and E$^n$ denotes 
the total energy at the n$^{\rm th}$ iteration}
\label{Fig:Cdiamond}
\end{figure}
\end{center}
We show the variation of the total energy as a function of the
number of electronic steps for the simple system using the tolerance 10$^{-10}$ Hartree in Figure~\ref{Fig:Cdiamond}.
In this example, the number of iterations 
\texttt{nline} carried out by the \textsc{lobpcg} or \textsc{pcg} method at each electronic step is set to 4, 
which is a typical value in \textsc{abinit} for many systems.   
Increasing \texttt{nline} does not change the number of electronic 
steps needed to reach the 10$^{-10}$ Hartree tolerance in this case and therefore is inefficient. 
Convergence is slightly faster for the \textsc{lobpcg} method, which takes 7 electronic 
steps vs. 10 steps needed by the \textsc{pcg}. The full-band \textsc{lobpcg} method does not 
improve the convergence of the SC loop in this case compared to the  band-by-band \textsc{lobpcg}.
\begin{center}
\begin{figure}[ht]
\mbox{\includegraphics[width=8.7cm]{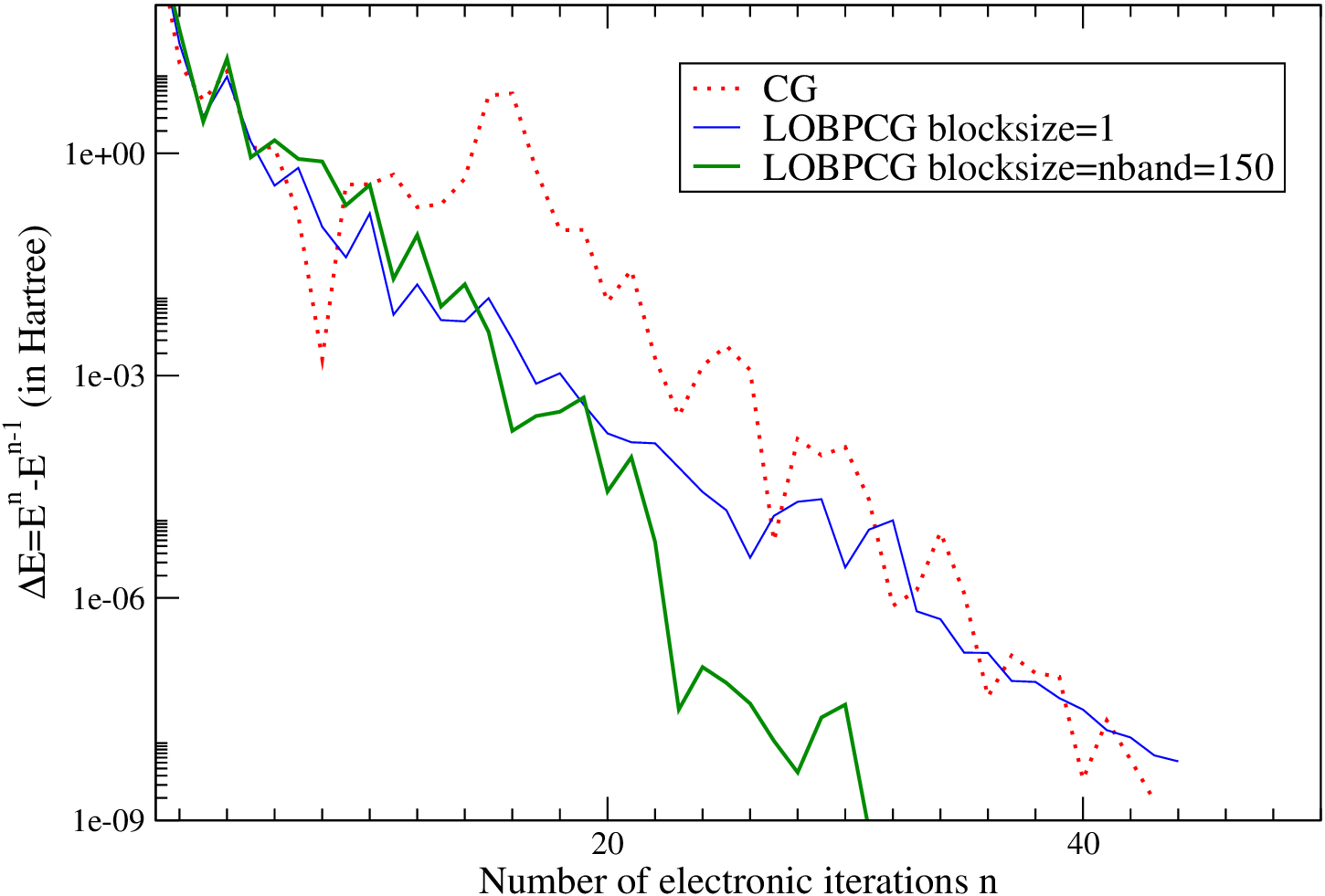}}
\caption{Same caption as Fig.~\ref{Fig:Cdiamond} for the plutonium in its $\alpha$ phase with 
$M=\texttt{nband}=150$.}
\label{Fig:Pualpha}
\end{figure}
\end{center}
For the plutonium in its $\alpha$ phase system, the typical value 
\texttt{nline}=4 appears too small, so we increase it to \texttt{nline}=6 
and display the results in Figure~\ref{Fig:Pualpha}.  
We observe that the number of electronic steps is 40 for both 
band-by-band \textsc{lobpcg} and \textsc{pcg} methods methods. But the full-band 
($m=\texttt{blocksize}=M=\texttt{nband}=150$)
\textsc{lobpcg} method finishes first with a clear lead in just 30 electronic steps. 
Further increase of \texttt{nline} does not noticeably affect the number of electronic steps. 

We conclude that the \textsc{lobpcg} method is competitive compared to the standard \textsc{pcg} method
and that choosing a larger block size $m>1$ in \textsc{lobpcg} may help to further improve the performance 
in sequential (one-processor) computations. 
However, the main advantage of the \textsc{lobpcg} method is that it can be efficiently 
parallelized as we describe in the next section. 
\section{The band-FFT parallelization}\label{sec:bfft}
\subsection{Distribution of the data}
The \textsc{lobpcg} algorithm works multiband-by-multiband, or block-by-block, 
so it can be easily parallelized over bands
by distributing the coefficients of the block over the processors. The main challenge 
is to make the multiband data partitioning compatible with the data partitioning
suitable for the 3D parallel FFT. In this section we describe these 
partitioning schemes and an efficient transformation between them. 

We implement two levels of parallelization using virtual Cartesian two dimensional MPI topology
with the so-called ``FFT-processors'' and ``band-processors'' along virtual horizontal and vertical directions, 
correspondingly.  
Let $\mathrm{P}$ be the number of planes along one direction of 
the 3D reciprocal lattice of vectors $\{\gras{G}\}$. 
We choose numbers $\mathrm{m}$ and $\mathrm{p}$ of band- and FFT-processors
to be divisors of $\mathrm{M}$ and $\mathrm{P}$, respectively, so that the total number of 
processors is \texttt{nproc}=$\mathrm{p\times m}$. 
All the quantities expanded over the PW basis set: wave-functions, densities, 
potentials, etc., undergo similar distributions. 
We focus below on the coefficients of the wave-functions, and make a comment about 
other data at the end of the section. 

We omit the index $\mathrm{k}$ in the following. We assume that the PW coefficients 
$\mathrm{c_n}(\gras{G})$ of a given band $\mathrm{n}$ are originally defined on the processor at the 
virtual corner 
and first the $\mathrm{P}$ planes are successively distributed over the $\mathrm{p}$ 
FFT-processors, such that the PW coefficients of all the bands are distributed along the 
first row of the virtual two dimensional processor topology:
\begin{displaymath}
 m \overbrace {\left\{
 \begin{array}{cccc}
  \mathrm{c_{n}}(\gras{G})&-&-&-\\
  -&-&-&-\\
  -&-&-&-\\
  -&-&-&-\\
 \end{array}
 \right\} }^p
\to 
 \left\{
 \begin{array}{cccc}
  \mathrm{c_{n}}(\gras{G}_{1})&\mathrm{c_{n}}(\gras{G}_{2})&\ldots&\mathrm{c_{n}}(\gras{G}_{\rm p})\\
  -&-&-&-\\
  -&-&-&-\\
  -&-&-&-\\
 \end{array}
 \right\}
\end{displaymath}
There are now $\mathrm{{P}/{p}}$ planes defined on each processor. The $\{\gras{G}_{\rm i}\}$ 
sub-sets for $\mathrm{i=1,2,\ldots,p}$ define a partition of the whole 3D 
reciprocal lattice of vectors $\{\gras{G}\}$. This distribution is typically used in conjunction
with Goedecker's parallel FFT routine.
To parallelize further, these planes are sub-divided again by distributing 
the PW coefficients over the $\mathrm{m}$ band-processors as:
\begin{displaymath}
 \left\{
 \begin{array}{cccc}
  \mathrm{c_{n}}(\gras{G}_{11})&\mathrm{c_{n}}(\gras{G}_{21})&\ldots&\mathrm{c_{n}}(\gras{G}_{\rm p1})\\
  \mathrm{c_{n}}(\gras{G}_{12})&\mathrm{c_{n}}(\gras{G}_{22})&\ldots&\mathrm{c_{n}}(\gras{G}_{\rm p2})\\
  \vdots&\vdots&\ddots&\vdots\\
  \mathrm{c_{n}}(\gras{G}_{\rm 1m})&\ldots&\ldots&\mathrm{c_{n}}(\gras{G}_{\rm pm})\\
 \end{array}
 \right\}
\end{displaymath}
The $\{\gras{G}_{\rm ij}\}$ sub-sub-sets for $\mathrm{j=1,2,\ldots,p}$ 
partition the  $\{\gras{G}_{\rm i}\}$ sub-set. 
When performing \textsc{lobpcg} calculations with blocks of size $\mathrm{m}$, 
the coefficients are grouped in $\mathrm{{{M}/{m}}}$ blocks of size $\mathrm{m}$,
e.g.,\ for the first $\mathrm{m}$ bands we have  
\begin{displaymath}
 \left\{
 \begin{array}{cccc}
  \mathrm{c_{1:m}(\gras{G}_{11})}&\mathrm{c_{1:m}(\gras{G}_{21})}&\ldots&\mathrm{c_{1:m}(\gras{G}_{p1})}\\
  \mathrm{c_{1:m}(\gras{G}_{12})}&\mathrm{c_{1:m}(\gras{G}_{22})}&\ldots&\mathrm{c_{1:m}(\gras{G}_{p2})}\\
  \vdots&\vdots&\ddots&\vdots\\
  \mathrm{c_{1:m}(\gras{G}_{1m})}&\ldots&\ldots&\mathrm{c_{1:m}(\gras{G}_{pm})}\\
 \end{array}
 \right\}
\end{displaymath}
This data distribution is well suited to perform dot-products as well as 
matrix-vector or matrix-matrix operations needed in the \textsc{lobpcg} algorithm. 
In particular, blocks of the Gram matrix are computed by the \textsc{blas} function \texttt{zgemm} 
on each processor using this distribution. In the next section we will show that the 
\texttt{zgemm} function plays the main role in superlinear scaling 
of the \textsc{lobpcg} algorithm. 
 
However, such a distribution is not appropriate to perform parallel 3D FFTs
and thus has to be modified. 
This is achieved by transposing the PW coefficients inside each column, so the following distribution 
of the first $\mathrm{m}$ bands is thus obtained:
\begin{displaymath}
 \left\{
 \begin{array}{cccc}
  \mathrm{c_{1}}(\gras{G}_{1})&\mathrm{c_{1}}(\gras{G}_{2})&\ldots&\mathrm{c_{1}}(\gras{G}_{\rm p})\\
  \mathrm{c_{2}}(\gras{G}_{1})&\mathrm{c_{2}}(\gras{G}_{2})&\ldots&\mathrm{c_{2}}(\gras{G}_{\rm p})\\
  \vdots&\vdots&\ddots&\vdots\\
  \mathrm{c_{m}}(\gras{G}_{1})&\ldots&\ldots&\mathrm{c_{m}}(\gras{G}_{\rm p})\\
 \end{array}
 \right\}
\end{displaymath}
where each band is distributed over the $\mathrm{p}$ FFT-processors.
The distribution is now suitable to perform 3D parallel FFTs. The efficient 3D FFT 
of Goedecker {\it et al.}~\cite{Goedecker_CPC_2003} implemented in our code is then applied on 
each virtual row of the processor partition, so $\mathrm{m}$ parallel 3D FFTs are performed 
simultaneously. 

The transformation used to transpose the coefficients within a column corresponds to the 
MPI\_ALLTOALL communication using the ``band communicator.'' During the 3D FFT, the MPI\_ALLTOALL
communication is also performed within each row using the ``FFT communicator.'' These two 
communications are not global, i.e., they do not involve communications between all processors,  
but rather they are local, involving processors only within virtual rows or columns. 
After \texttt{nline} \textsc{lobpcg} iterations for this first block of bands, we apply the 
same strategy to the next block, with the constraint that all bands in the block are 
orthogonal to the bands previously computed.

Finally, this last distribution is used to compute by FFT the contributions of
a given band to the density, which are subsequently summed up over
all bands-processors to calculate the electronic density distributed over
the FFT-processors. The electronic density is in its turn used to produce
local potentials having the same data distribution.
\subsection{Subspace diagonalization\label{sec:sd}}
Here we address the issue (ii) brought up in section \ref{sec:SC} that  
diagonalization of a matrix of the size $M\gg1$ is computationally expensive. The standard 
implementation of this calculation is performed in the sequential mode by calling \textsc{lapack}
to diagonalize the same matrix on every processor. It becomes inefficient if the number of bands $M$ 
or the number of processors increases. 
To remove this bottleneck, we call the \textsc{scalapack} library, which is a 
parallel version of \textsc{lapack}. The tests performed using \textsc{scalapack} show that 
this bottleneck is essentially removed for systems up to 2000 bands.
For instance, on 216 processors this diagonalization takes 23\% of the total time for a system 
with 1296 bands using  \textsc{lapack} on each processor, but only 3\% using \textsc{scalapack}. 
\subsection{Generalization of the transposition principle\label{sec:tp}}
Finally, we address bottleneck (iv) formulated in section \ref{sec:SC}, i.e.,\   
computation of the three non-local like terms (the non-local part of the potential v$_{\rm nl}$, 
the overlap operator $\mathcal{O}$ and the $\rho_{\rm ij}$ matrix), which is one of the most time 
consuming parts of the SC loop. Parallelization is straightforward if these terms are computed 
in reciprocal space and the distribution of the PW is efficient. The data distribution 
here corresponds to the one used in the \textsc{lobpcg}. 

Computation of these terms can be made more efficient in parallel if we utilize
the data distribution used for the 3D parallel FFTs based on one virtual 
line of processors \cite{Hutter_CPC6_2005,Hutter_PC31_2005}. 
The cause of better efficiency here comes from merging 
two parallel features. On the one hand, we avoid any collective communication on the whole set 
of processors, thus reducing the data over each line independently rather than over the whole set of processors. 
On the other hand, we obtain better load-balancing of the processors since
the FFT data are not sub-divided in this case.
\section{Numerical Results\label{sec:results}}
Calculations are performed on Tera-10 NovaScale 5160 supercomputer,
composed of 544 nodes (with 8 dual-core 1.6 GHz Intel Montecito processors per node) connected by Quadrics  
at the ``Comissariat \`a l'\'energie atomique.'' 
Benchmarks are carried out on a supercell calculation with 108 atoms of gold set in 
their Face Centered Cubic lattice positions and 648 bands. An energy cutoff of 24 Ha is used in these 
calculations, leading to a three dimensional reciprocal grid with 108$^3$ points. 
In all tests \texttt{nline}=4.  
\subsection{The two-level parallelization}
Calculations are performed on various numbers of processors \texttt{nproc}=1, 4, 18, 54, 
108, 162, and 216. For each value of \texttt{nproc}, we consider all distributions 
$\mathrm{m\times p}$ which are allowed within this framework.  
We remind the reader that $\mathrm{p}$ has to be smaller than 108/2=54, 
in order to use the 3D parallel FFT, whereas $\mathrm{m}$ and $\mathrm{p}$ have to be divisors of  
$\mathrm{M=648}$ and $\mathrm{P=108}$, respectively. For instance, for \texttt{nproc}=18, 
we can choose $\mathrm{m\times p}=$18$\times$1, 
9$\times$2, 6$\times$3, 3$\times$6, 2$\times$9 and 1$\times$18.

\begin{center}
\begin{figure}[ht]
\mbox{\includegraphics[width=8.7cm]{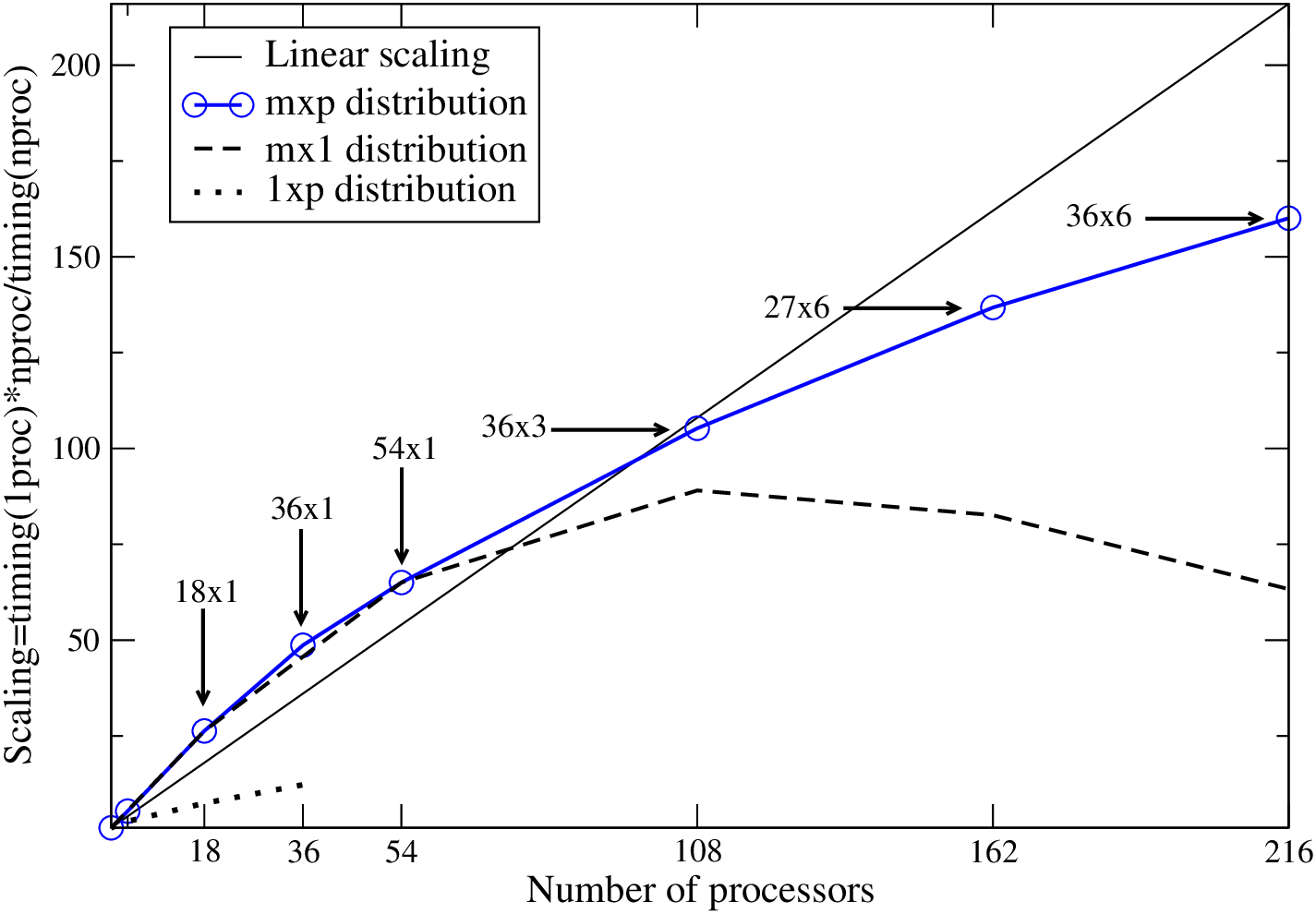}}
\caption{Two-level parallelization: Speedup of the \textsc{abinit} code with respect to the number of 
the processors for  
$\mathrm{m\times p, m\times 1}$, and 1$\times\mathrm{p}$ data distributions. 
The ``ideal'' linear speedup is displayed by a line with the slope 1.  
For each arrow, we give numbers m and p of band- and FFT-processors, correspondingly.}
\label{Fig:scaling-2D}
\end{figure}
\end{center}

We focus on three kinds of benchmarks here: the first two are $\mathrm{m}\times$1 (band-only)  
and 1$\times\mathrm{p}$ (FFT-only) distributions. 
The virtual MPI topology is one dimensional in these cases. 
The third kind is the optimized m$\times$p (band-FFT) distribution.
The speedups for these three parallelization schemes are shown in Figure~\ref{Fig:scaling-2D}.
The transposition of the data distribution within each column during the computation of $\mathrm{v}_{\rm nl}$ and the use of collective 
communications inside lines for the FFT, applying $\mathrm{v}_{\rm loc}$ to the wave-functions, are
the two most important communications used in the framework of the Band-FFT parallelization. The 
communications within the FFTs increase to 17\% of the total time on 216 processors 
using the 1$\times\mathrm{p}$ (FFT-only).
In contrast, the $\mathrm{m}\times$1 (band-only) distribution  
is adequate for up to 54 processors. However, further increase 
of the block size $\mathrm{m}$ together with the number of processors 
is inefficient in this case for more than 100 processors. We remind the reader that the \textsc{lobpcg}
solver performs the Rayleigh-Ritz procedure on subspaces of dimension 
3$\mathrm{m}$ on every iteration. For large $\mathrm{m}$ this starts 
dominating the computational costs even on $\mathrm{m}$ processors.  
So we reduce $\mathrm{m}$ and introduce our double parallelization.
For the optimal band-FFT parallelization we obtain a superlinear scaling up to 100 
processors, and a speedup of 150 for 200 processors. 
\begin{center}
\begin{figure}[ht]
\mbox{\includegraphics[width=8.7cm]{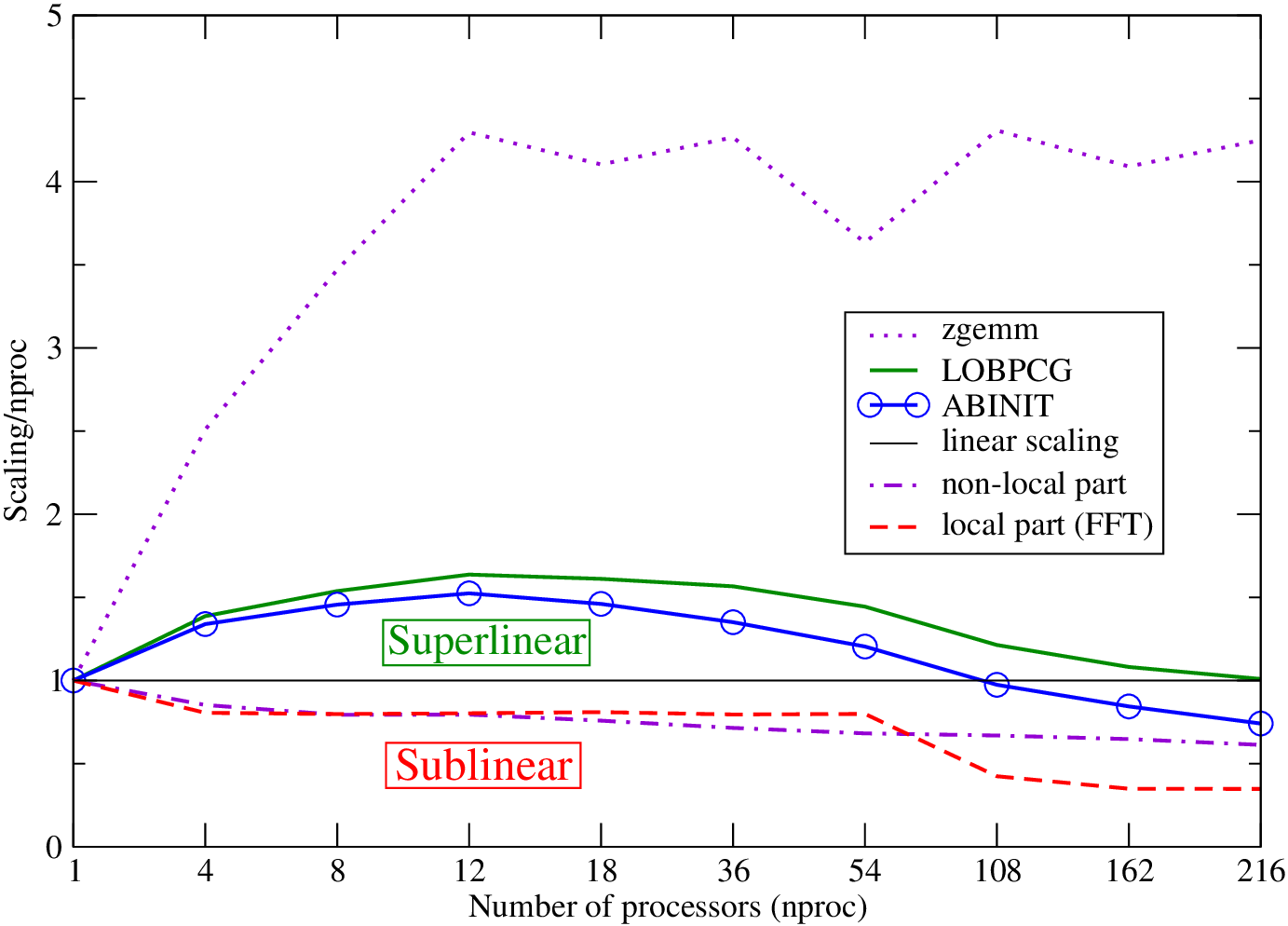}}
\caption{Two-level parallelization: Speedup of various parts of the SC loop using the 
same $\mathrm{m\times p}$ distribution as the best (top) curve in  Figure~\ref{Fig:scaling-2D}. 
The main parts displayed are: \texttt{zgemm}, non-local ($\mathrm{v}_{\rm nl}$) 
 and local ($\mathrm{v}_{\rm loc}$ using FFT) parts of the potential. The \textsc{lobpcg} 
represents mainly the sum of the above. The \textsc{abinit} curve is the same as on Figure~\ref{Fig:scaling-2D}.}
\label{Fig:bandfft-publi}
\end{figure}
\end{center}
Figure~\ref{Fig:bandfft-publi} displays scaling of different parts of the \textsc{abinit} code 
using a slightly different format: the vertical axis represents the scalability divided 
by the number of processors, so the linear scaling is now 
displayed by the horizontal line. We also zoom in the horizontal axis to better 
represent the scalability for small numbers of processors. The \textsc{abinit} curve 
shows exactly the same data as in Figure~\ref{Fig:scaling-2D}. The \textsc{lobpcg} curve 
represents the cumulative timing for the main parts that the \textsc{lobpcg} function calls  
in \textsc{abinit}: \texttt{zgemm}, non-local ($\mathrm{v}_{\rm nl}$) 
 and local ($\mathrm{v}_{\rm loc}$ using FFT) parts of the potential. 
 
Figure~\ref{Fig:bandfft-publi} demonstrates that the super-linear behavior 
is due to the \texttt{zgemm}. In the sequential mode we run the 
band-by-band eigensolver that does not take much advantage of the \textsc{BLAS} function
\texttt{zgemm} use, since only matrix-vector products are performed in this case.  
So the super-linear behavior is not directly related to the number of processors, 
but is rather explained by the known positive effect of vector blocking in \textsc{lobpcg} that 
allows calling more efficient \textsc{BLAS} level 3 functions for matrix-matrix operation
rather then \textsc{BLAS} level 2 functions for matrix-vector operations. 
This effect is especially noticeable in \textsc{lobpcg} since it is intentionally written 
in such a way that the vast majority of linear algebra related operations are 
matrix-matrix operations (performed in \textsc{lobpcg} using the \texttt{zgemm}),
which represent 60\% of the total time in band-by-band calculations on a single processor,  
but exhibits a strong speedup with the increase of the block size in the parallel mode. 

The acceleration effect is hardware-dependent and is a result on such subtle processor 
features as multi-level cache, pipelining, and internal parallelism.  
Using the hardware-optimized \textsc{blas} is crucial: in one of our tests 
the time spent in the \texttt{zgemm} routine represents only 10\% of the total time 
using block size $m=\texttt{nproc}=54$ if hardware-optimized \textsc{blas} is used, but 53\% otherwise. 
All the computation results reported here are obtained by running the \textsc{abinit} code 
with the hardware-optimized \textsc{blas}. 

The effectiveness of the \textsc{blas} level 3 matrix-matrix \texttt{zgemm} also 
depends on the sizes of the matrices involved in the computations. 
This is the point where the parallel implementation that partitions the matrices 
may enter the scene and play its role in the acceleration effect. 
%
\subsection{The three-level parallelization}
Parallelization over \textbf{k}-points is commonly used in {\it ab initio} calculations and is 
also available in \textsc{abinit}. 
It is very efficient and easy to implement, and results in a roughly linear scaling 
with respect to the number of processors. 
This kind of parallelization is evidently limited by the number $\mathrm{k}$ of \textbf{k}-points available in 
the system. It appears theoretically useless for large supercell calculations or disorder systems with 
only the $\Gamma$ point within the Brillouin zone, but 
in practice sampling of the Brillouin zone is sometimes needed, e.g.,\
for surfaces and interfaces in the
direction perpendicular to the stacking sequence~\cite{Kresse_PRB68_2003,Segeva_SS601_2007}, or 
for accurate thermodynamic integrations for metling~\cite{Cricchio_PRB73_2006,Wang_PRL95_2005}.
Parallelization over \textbf{k}-points is clearly well suited for metals where 
a large number $\mathrm{k}$ of \textbf{k}-points is required. 

We now combine the \textbf{k}-point and the Band-FFT parallelization by constructing 
a virtual three-dimensional configuration 
of processors $\mathrm{m\times p\times k}$, where the Band-FFT parallelization
is applied on each \textbf{k}-point in parallel.
\begin{center}
\begin{figure}[ht]
\mbox{\includegraphics[width=8.7cm]{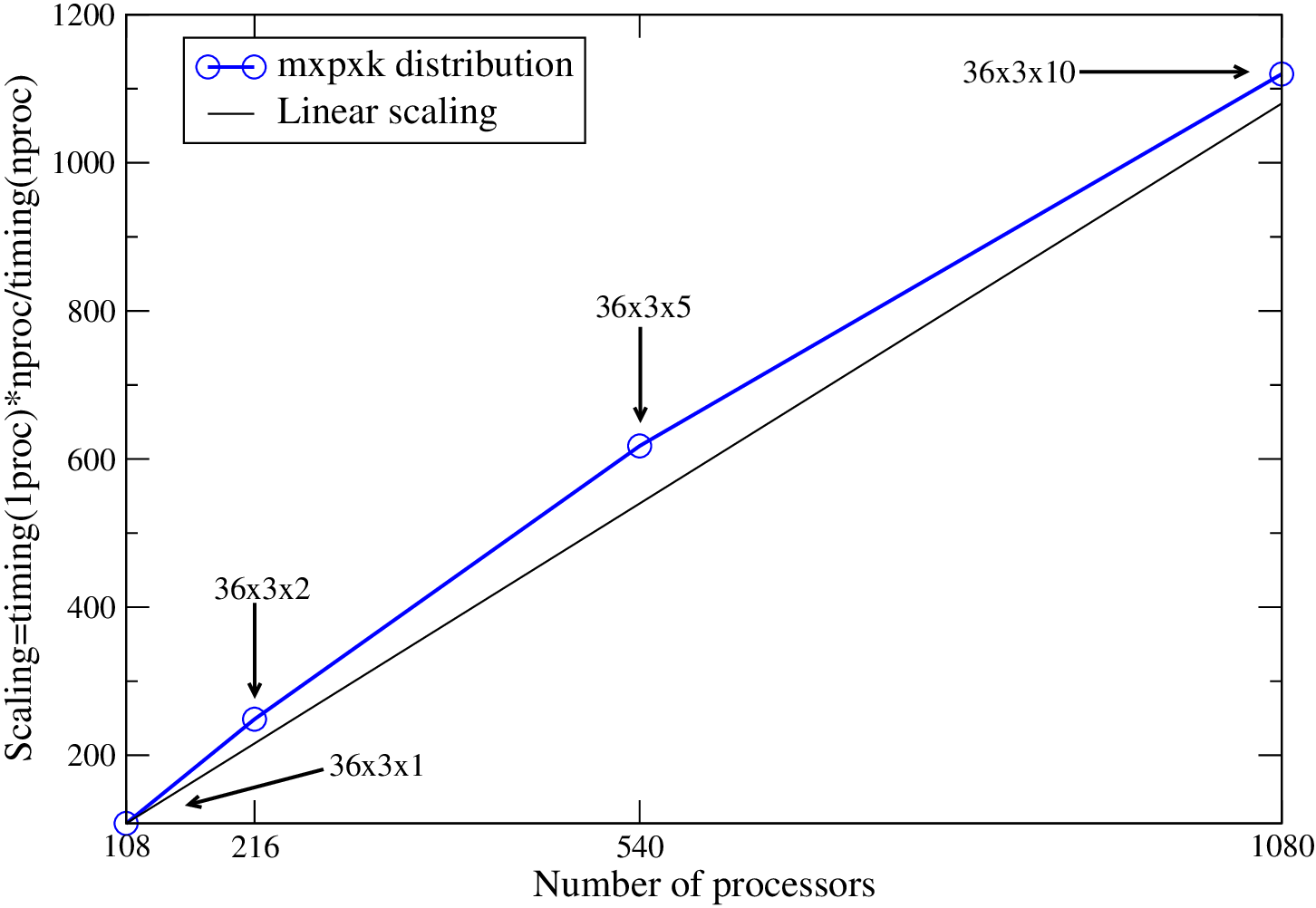}}
\caption{The triple parallelization: speedup of the \textsc{abinit} code with respect to the number 
of processors.}
\label{Fig:scaling-3D}
\end{figure}
\end{center}
In Figure~\ref{Fig:scaling-3D} we present the speedup obtained for this triple parallelization. 
We use the same system as previously except that 10 \textbf{k}-points
rather than 1 are sampled in the Brillouin zone.  
We use as the reference the limit of the super-linear scaling obtained in the framework of the 
Band-FFT parallelization: 108 processors with a 36$\times$3 distribution. As expected,
adding \textbf{k}-point processors in the virtual third dimension of the 3D 
processor configuration results in the linear scaling of computational costs. 
Using our three-level parallelization on 10 \textbf{k}-points  
the \textsc{abinit} scales linearly up to 1000 processors. 
\section{Conclusion}
Our combined \textbf{k}-points-multiband-FFT parallelization allows 
the material science community to 
perform large scale electronic structure calculations efficiently  
on massively parallel supercomputers up to thousand of processors using 
the publicly available software \textsc{abinit}.  
The next generation of petascale supercomputers will bring new challenges. 
Novel numerical algorithms and advances in computer sciences, e.g.,  
introduction of asynchronous (non-blocking) 
collective communication standards~\cite{hoefler-standard-nbcoll}, would help us 
to minimize the communication load in future versions of \textsc{abinit} 
and make efficient use of the next generation hardware.  
\begin{ack}
The authors thank A. Curioni for fruitful 
discussions about the double parallelization technique and T. Hoefler about 3D FFT in parallel.
\end{ack}

\end{document}